%% file: Ay_QE_PRL.tex
\documentclass[reprint,superscriptaddress,showpacs,preprintnumbers,amsmath,amssymb]{revtex4-1}
\usepackage{placeins}
\usepackage[usenames]{color}
\usepackage{graphicx}
\usepackage{dcolumn}
\usepackage{bm}

\begin{document}

\title{Measurement of the Target-Normal Single-Spin Asymmetry in Quasi-Elastic Scattering from the Reaction $^{3}$He$^{\uparrow}$($e,e^\prime$)}

\input author_list.tex
\date{\today}

\begin{abstract}
We report the first measurement of the  target single-spin asymmetry, $A_y$, in quasi-elastic scattering from 
the inclusive reaction $^3$He$^{\uparrow}\left(e,e^\prime \right)$  on a $^3$He gas target 
polarized normal to the lepton  scattering plane. Assuming time-reversal invariance, this asymmetry is strictly zero for one-photon exchange.   A non-zero $A_y$ can arise from the interference between the one- and two-photon exchange processes which is sensitive to the details of the sub-structure of the nucleon.  An experiment recently completed at Jefferson Lab  yielded asymmetries with high statistical precision  at $Q^{2}=$ 0.13, 0.46 and 0.97 GeV$^{2}$.   These measurements demonstrate, for the first time, that the $^3$He asymmetry is clearly  non-zero and negative with a statistical significance of (8-10)$\sigma$. Using measured proton-to-$^{3}$He cross-section ratios and the effective polarization approximation, neutron asymmetries  of $-$(1-3)\% were obtained.  The neutron asymmetry at high $Q^2$ is related to moments of the Generalized Parton Distributions (GPDs).  Our measured neutron asymmetry at $Q^2=0.97$ GeV$^2$ agrees well with a prediction based on two-photon exchange using a GPD model and thus provides a new, independent constraint on these distributions.
\end{abstract}

\maketitle

Elastic and inelastic form factors, extracted from electron-nucleon scattering data, provide invaluable information on nucleon structure.  In most cases the scattering cross sections are dominated by one-photon exchange. Contributions from two-photon exchange are suppressed relative to the one-photon exchange contribution but are important in certain processes.   

\begin{figure}[h]
\includegraphics[width=3in]{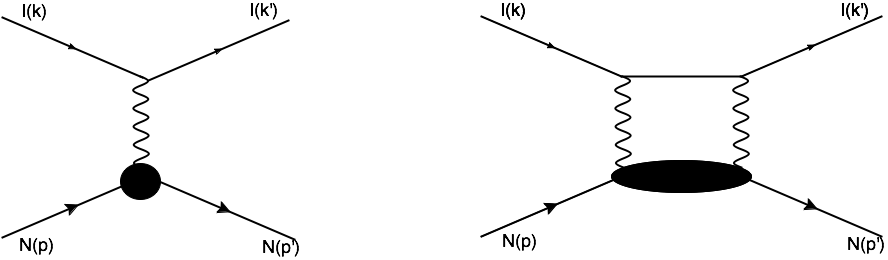}
\caption{In inclusive electron scattering  a non-zero target-normal SSA can arise due to interference between one- (left) and two- (right) photon exchange.  Here $N$ is the nucleon with incident and outgoing 4-momenta $p$ and $p^\prime$, respectively, and $l$ is the lepton  with incident and outgoing 4-momenta $k$ and $k^\prime$, respectively.  The  intermediate nucleon state, represented by the black oval, includes both elastic and inelastic contributions and is thus sensitive to the structure of the nucleon.} 
\label{fig:TPEX_feynman} 
\end{figure}

One observable that is exactly zero for one-photon exchange is the target-normal single-spin asymmetry (SSA), $A_y$, which is the focus of this experiment.   When two-photon exchange is included, $A_y$ can be non-zero.  As shown in Fig.~\ref{fig:TPEX_feynman} the two photons form a loop that contains the nucleon 
intermediate state which has an elastic contribution that is calculable~\cite{Chen:2004tw},  and an inelastic contribution that must be modeled.  This makes the two-photon exchange process sensitive to the details of nucleon structure and provides a powerful new tool for testing model predictions.  

Recently, $A_y$ for the neutron ($^3$He) was measured to be non-zero and negative at the 2.89$\sigma$ level for deep-inelastic scattering~\cite{Katich:2013atq}.  Two-photon-exchange contributions are also important  when extracting the proton elastic form factor $G_E^p(Q^2)$ from measured data.   Values extracted from Rosenbluth separation of cross section data differ markedly from those extracted from polarization-transfer measurements~\cite{g63, Arrington:2011dn, Puckett:2011xg, Blunden_GeGm,  Chen:2004tw, carl_2gamma}.  A GPD-based model prediction for the two-photon exchange contributions reduced the discrepancy by $\sim50\%$ for $Q^2\geq 1 $GeV$^2$~\cite{Chen:2004tw}.   This same model also  predicts $A_y\sim-2\%$ at $Q^2=1$ GeV$^2$. A measurement of $A_y$  is thus an independent test of the GPD model in the absence of a large  contribution from one-photon exchange.  

\begin{figure}[h]
\includegraphics[width=3in]{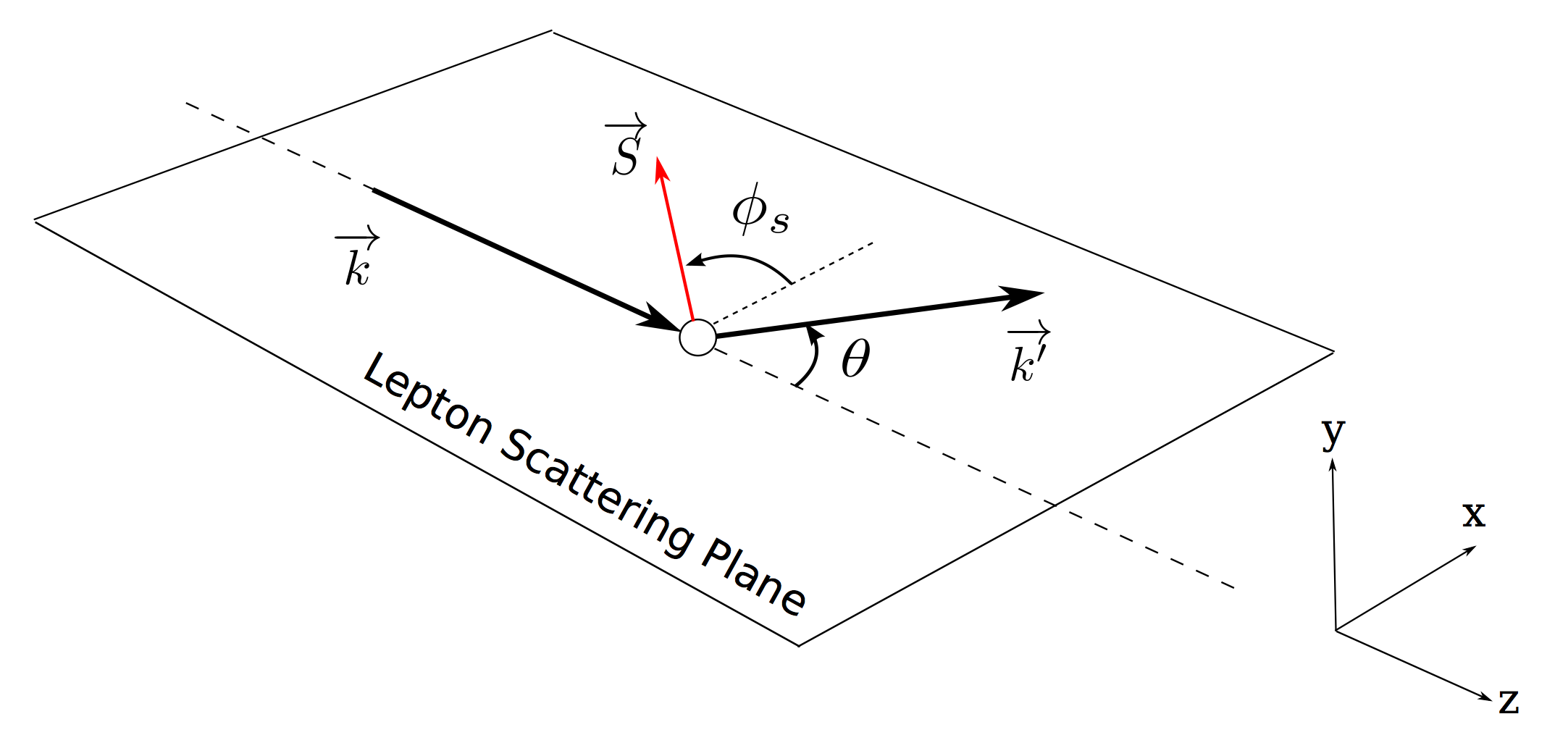}
\caption{Coordinate system used to define $A_{UT}(\phi_S)$.} 
\label{fig:coords} 
\end{figure}

Consider the elastic scattering of an unpolarized electron from a target nucleon with  
spin $\vec{S}$, oriented perpendicular (transversely polarized) to the incident electron 3-momentum $\vec{k}$, and normalized such that $|\vec{S}|=1$.   Requiring conservation of both the electromagnetic current 
and parity, the differential cross section, $d\sigma$, for the inclusive $(e,e^\prime)$ reaction is 
written as~\cite{Christ_Lee,Cahn_2g,Afanasev_DIS} 
\begin{eqnarray}
d\sigma(\phi_S)  &=& d\sigma_{UU} + \frac{{\vec{S}}\cdot({\vec{k}} \times {\vec{k'}})}{|{\vec{k}} \times {\vec{k'}}|}   d\sigma_{UT}\nonumber \\
&=& d\sigma_{UU} + d\sigma_{UT} \sin \phi_S, 
\end{eqnarray}
where $\vec{k'}$ is the  3-momentum of the scattered electron, and  $d\sigma_{UU}$ and $d\sigma_{UT}$ are the cross sections for an unpolarized electron scattered from an unpolarized and transversely polarized target, respectively.   
Our choice of coordinates is shown in Fig.~\ref{fig:coords} with the angle $\phi_S$ between the lepton plane and $\vec{S}$.  The $+\hat{y}$ direction is parallel to the vector ${\vec{k}} \times {\vec{k'}}$ and corresponds to $\phi_S=90^{\circ}$. 
We define the target SSA as
\begin{equation}\label{eq:ay}
A_{UT} (\phi_S) = \frac{d\sigma(\phi_S)-d\sigma(\phi_S + \pi)}{d\sigma(\phi_S)+d\sigma(\phi_S + \pi)}
 =  A_y \sin \phi_S.
\end{equation}
By measuring   $A_{UT}$ at $\phi_S=\pi/2$  one can extract the quantity $A_y \equiv \frac{d\sigma_{UT}}{d\sigma_{UU}}$, which is the SSA for a target polarized normal to the lepton plane.

For one-photon exchange we can write 
$d\sigma_{UU} \propto \mathrm{Re}(\mathcal{M}_{1\gamma} \mathcal{M}_{1\gamma}^*)$ and 
$d\sigma_{UT}\propto \mathrm{Im}(\mathcal{M}_{1\gamma} \mathcal{M}_{1\gamma}^*)$,  where $\mathcal{M}_{1\gamma}$ is the one-photon exchange amplitude and $\mathrm{Re}$ ($\mathrm{Im}$) stands for the real (imaginary) part.  However, time-reversal 
invariance  requires that $\mathcal{M}_{1\gamma}$ be real and so at order $\alpha^2$, $d\sigma_{UU}$ 
can be non-zero, but $d\sigma_{UT}$ must be zero~\cite{Christ_Lee}.  When one includes the (complex) two-photon exchange 
amplitude, $\mathcal{M}_{2\gamma}$, the contribution to  the asymmetry from  the interference between one- and two-photon 
exchange amplitudes is $d\sigma_{UT}\propto \mathrm{Im}(\mathcal{M}_{1\gamma} \mathcal{M}_{2\gamma}^*)$ 
which can be non-zero at order $\alpha^3$.

Using the formalism of  Ref.~\cite{Chen:2004tw}, we can write
\begin{eqnarray}
						  					\label{eq:tnsa}
A_y &=& \sqrt{\frac{2 \, \varepsilon \, (1+\varepsilon )}{\tau}} \,\,
	\frac{1}{\sigma_R} 
	\left\{ - \, G_M \, \mathrm{Im} 
	\left(\delta \tilde G_E + \frac{\nu}{M^2} \tilde F_3 \right) \right.\nonumber \\
	&+ &\, \left. G_E \, \mathrm{Im} \left(\delta \tilde G_M 
	+ \left( \frac{2 \varepsilon}{1 + \varepsilon} \right) 
	\frac{\nu}{M^2} \tilde F_3 \right) \right\} 
	       \, ,  
\end{eqnarray}
where  $\tau \equiv Q^2/4M^2$,  $\nu=\frac{1}{4}(k_\mu +k^\prime_\mu)(p^\mu +p^{\prime\mu}) $,  $\varepsilon \equiv (1 + 2(1+\tau) \tan^2 {\theta\over 2})^{-1}$ and $M$ is the mass of the nucleon.  
In the lab  frame, $E$, $E'$  and $\theta$ are the incident and scattered energies, and scattering angle, of the electron, respectively.   The $G_E$ and $G_M$ are the Sachs form factors and $\sigma_R$ is the unpolarized cross section.
The terms $\delta \tilde G_E$, $\delta \tilde G_M$ and  $\tilde F_3$ are additional complex contributions that arise when two-photon exchange is included. They are exactly zero for one-photon exchange.  For the neutron, unlike the proton, $G_E<<G_M$ so that Eq.~(\ref{eq:tnsa}) is dominated by the term proportional to $G_M$.  Note that the unpolarized cross section  and polarization transfer observables depend on the real parts of $\delta \tilde G_E$, $\delta \tilde G_M$ and $\tilde F_3$.

For $Q^2\geq 1$ GeV$^2$ the two-photon contributions to Eq.~(\ref{eq:tnsa}) were estimated using weighted moments of the GPDs,  $H^q$, $E^q$ and $\tilde{H}^q$, for a quark $q$~\cite{Chen:2004tw}.   For lower $Q^2$, $A_y$ can be estimated using  model fits of nucleon resonance and pion production data~\cite{Blunden_GeGm, Pasquini:2004pv}. However, there are no predictions in the kinematic range of this experiment.   The only existing measurement was made on the proton at SLAC in 1970~\cite{Powell:1970qt}.  They measured asymmetries at $Q^2=0.38$, $0.59$ and $0.98$ GeV$^2$ that were consistent with zero at the $\sim10^{-2}$ level.  There has never been a corresponding measurement  on the neutron.
 
This paper presents the results of Jefferson Lab experiment E05-015, which measured $A_{y}$ by scattering unpolarized electrons from $^3$He nuclei polarized normal to the electron scattering plane.  The electron beam was longitudinally polarized with energies of 1.2, 2.4 and 3.6 GeV and an average current of 12 $\mu$A (CW).  The helicity of the beam was flipped at a rate of 30 Hz (for other experiments requiring a polarized electron beam) and data from the two helicity states were summed for this analysis.

The polarized target used in this experiment was a 40 cm-long aluminosilicate glass cell filled with  $^3$He gas at a density of 10.9 amg.  A small quantity of N$_2$ gas, $\sim 0.1$ amg, was also included to aid in the polarization process. The target was polarized through spin-exchange optical pumping of a Rb-K mixture~\cite{Singh:2013nja}. In order to reduce the systematic uncertainty, the direction of the target polarization vector was  reversed every 20 minutes using adiabatic fast passage. The polarization was monitored during each spin-flip using nuclear magnetic resonance. Electron paramagnetic resonance  measurements were periodically made throughout the experiment in order to calibrate the polarization~\cite{Romalis_EPR}. The  average in-beam target polarization was (51.4$\pm$2.9)$\%$. 

The electron beam was rastered in a 3 mm $\times$ 3 mm pattern to reduce the possibility of cell rupture due to localized heating of the thin glass windows.   Electrons scattered from the target were detected using the two Hall A high resolution spectrometers (HRSes)~\cite{Alcorn:2004sb} at scattering angles of $\pm 17^{\circ}$ with respect to the incident beam direction.  
Both spectrometers were configured to detect electrons in single-arm mode using nearly identical detector packages, each consisting of two dual-plane vertical drift chambers for tracking, two planes of segmented plastic scintillator for trigger formation, and  CO$_{2}$ gas Cherenkov  and Pb-glass electromagnetic calorimeter detectors for hadron rejection. The data acquisition systems for the spectrometers were synchronized   to allow cross-checking of the results.  By simultaneously measuring with two independent spectrometers, we confirmed that the measured asymmetries were consistent in magnitude, with opposite signs, as expected.

Consistent with Ref.~\cite{Katich:2013atq},  the lepton scattering plane is defined by the incoming and outgoing electron momenta, $\vec{k}$ and $\vec{k'}$, respectively, as in Figure~\ref{fig:coords}. If the target spin is parallel to $\hat{y}$, we define it as spin up ($\uparrow$), while the target spin is anti-parallel to $\hat{y}$ is defined as spin down ($\downarrow$).

The electron yields, $Y^{\uparrow(\downarrow)}$, give the number of electrons ($N^{\uparrow(\downarrow)}$) in the target spin-up (spin-down) state that pass all the particle-identification cuts, normalized by accumulated charge ($Q^{\uparrow(\downarrow)}$) and data-acquisition live-time ($LT^{\uparrow(\downarrow)}$):
\begin{equation}
Y^{\uparrow(\downarrow)} = \frac{N^{\uparrow(\downarrow)}}{Q^{\uparrow(\downarrow)}\cdot LT^{\uparrow(\downarrow)}}
\label{eqn:Y}
\end{equation}

The raw experimental asymmetries were calculated as
\begin{equation}
A_\mathrm{raw} = \frac{Y^{\uparrow} - Y^{\downarrow}}{Y^{\uparrow} + Y^{\downarrow}}
\label{eqn:A_raw}
\end{equation}
and were corrected for nitrogen dilution and target polarization. The nitrogen dilution factor is defined as
\begin{equation}
f_\mathrm{N_{2}}\equiv\frac{\rho_\mathrm{N_{2}}\sigma_\mathrm{N_{2}}}{\rho_\mathrm{^{3}He}\sigma_\mathrm{^{3}He}+\rho_\mathrm{N_{2}}\sigma_\mathrm{N_{2}}},
\label{eqn:fn}
\end{equation}
where $\rho_i$ and $\sigma_i$ are the number densities and unpolarized cross sections, respectively. The nitrogen density was measured when filling the target cell and the cross-section was determined experimentally by electrons scattering from a reference cell filled with a known quantity of N$_{2}$.  The denominator was obtained from the polarized target cell yields.

\begin{figure}
\includegraphics[width=\linewidth]{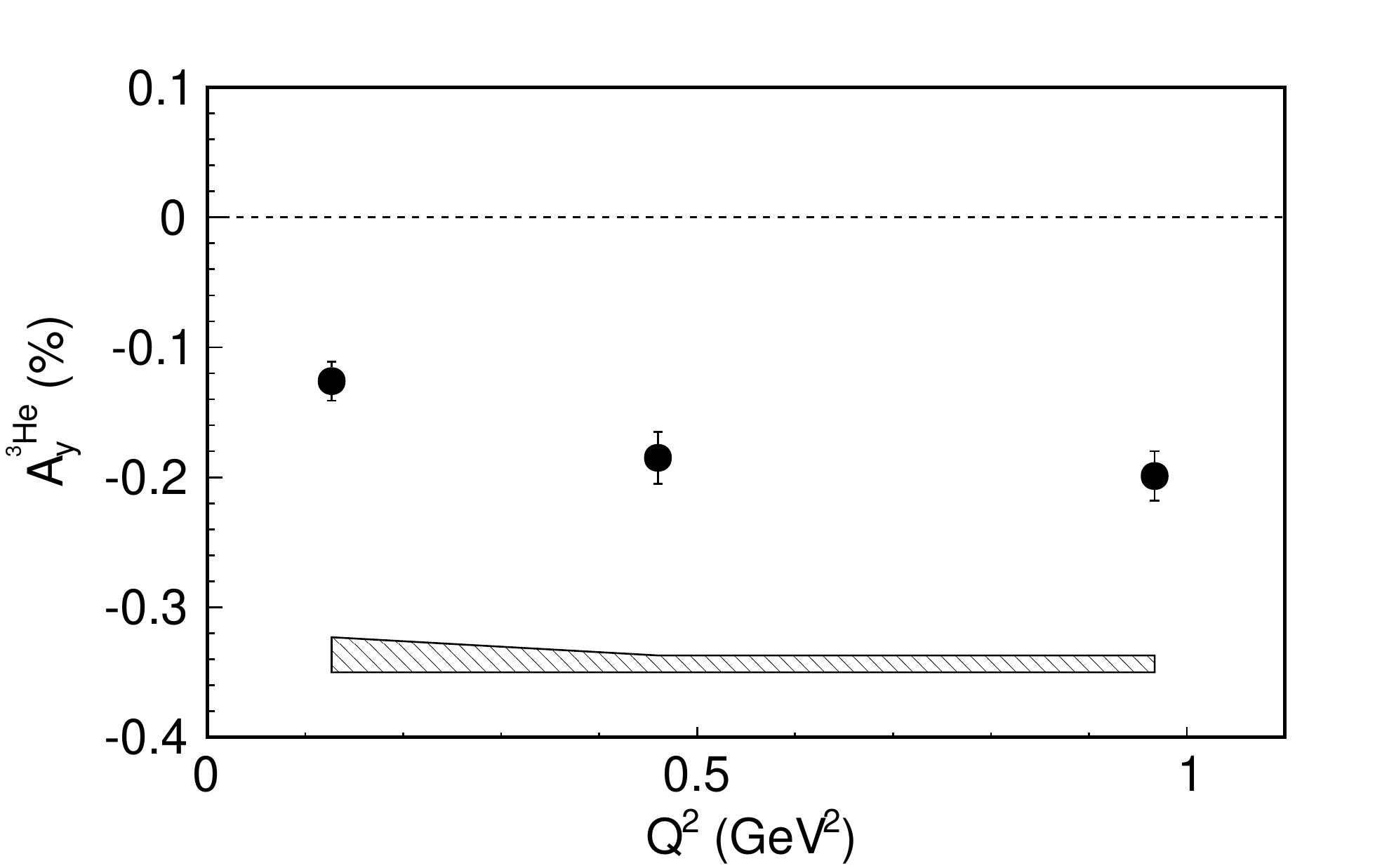}
\caption{\label{fig:result_he3_q2} Measured $^\mathrm{3}$He asymmetries, $A_{y}^\mathrm{^{3}He}$, as a function of $Q^2$.  Uncertainties shown on the data points are statistical.  Systematic uncertainties are shown by the band at the bottom.}
\end{figure}

The final asymmetries were obtained after subtraction of the elastic
radiative tail contribution, radiative corrections of the quasi-elastic
asymmetries and corrections for bin-averaging effects. The contribution of the elastic
radiative tail to the lowest $Q^2$ point  was 3\%, and was negligible for the two larger $Q^2$ points.  Results for $A_{y}^\mathrm{^{3}He}$ are shown in Fig.~\ref{fig:result_he3_q2} and listed in Table~\ref{tab:results_n}. The uncertainties on the data points are statistical, with the total experimental systematic uncertainty indicated as an error band below the data points. The systematic uncertainty in $A_{y}^\mathrm{^{3}He}$ includes contributions from the live-time asymmetry, target polarization, target misalignment, nitrogen dilution and radiative corrections.   The dominant contribution to the systematic uncertainty at the two largest $Q^2$ points is the uncertainty in the target polarization,  $\pm5.6\%$ (rel.).  At the two largest $Q^2$ points, the results from the left and right HRS agree to  $<1\sigma$ (stat.).  At the lowest $Q^2$ point the results differ by $\sim 2\sigma$ (stat.), which we included in the systematic uncertainty.

Polarized $^3$He targets have been used in many experiments as  an effective polarized neutron target~\cite{Kuhn:2008sy,Qian:2011py}. The ground state of the $^3$He nucleus is dominated by the $S$-state in which the two proton spins are  anti-parallel, and the nuclear spin is carried by the neutron~\cite{PhysRevC.65.064317}.  From the polarized $^3$He asymmetries, the neutron asymmetries, $A_{y}^{n}$, were extracted  using the effective neutron polarization approximation~\cite{Scopetta}, 

\begin{equation}
A_{y}^{n}=\frac{1}{(1-f_{p})P_{n}}(A_{y}^\mathrm{^{3}He}-f_{p}P_{p}A_{y}^{p})
\label{eqn:pdilution}
\end{equation}
where the proton dilution factors, $f_{p}$, were  calculated using the Kelly parametrizations for $G_E^p$ and $G_M^p$~\cite{PhysRevC.70.068202}  and assuming $f_{p}=(1-f_n)=2\sigma_{p}/\sigma_\mathrm{^{3}He}=2\sigma_{p}/(2\sigma_p+\sigma_{n})$. At the lowest two $Q^2$ points, where nuclear effects may be important,  a calculation was also made by A.~Deltuva~\cite{PhysRevC.68.024005, PhysRevC.69.034004, PhysRevC.70.034004, PhysRevC.72.054004} using a non-relativistic model of the $^3$He nucleus based on the CD-Bonn potential. Results extracted using the Kelly and Deltuva methods are consistent.

\begin{figure}
\includegraphics[width=\linewidth]{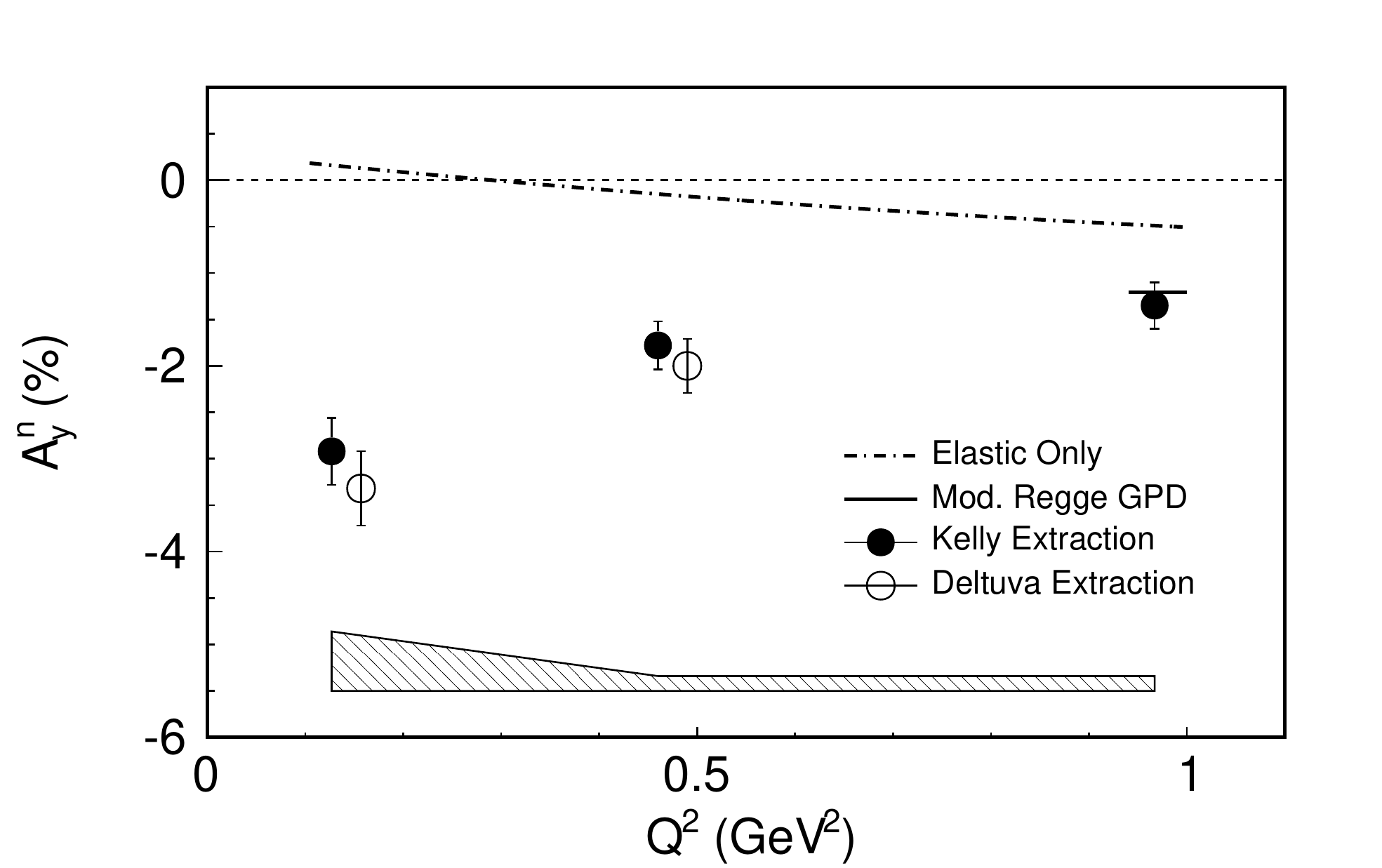}
\caption{\label{fig:result_n_q2} Results for the neutron asymmetries, $A_y^\mathrm{n}$, as a function of $Q^2$.  Uncertainties shown on the data points are statistical.  Systematic uncertainties are shown by the band at the bottom.  The elastic contribution to the intermediate state is shown by the dot-dash line~\cite{Afanasev:2002gr},  and at $Q^2=0.97$ GeV$^2$, the GPD calculation of Chen {\em{et al.}}~\cite{Chen:2004tw} is shown by the short solid line.}
\end{figure}

The effective neutron and proton polarizations in $^{3}$He are given by P$_{n}$=0.86$_{-0.02}^{+0.036}$ and P$_{p}$=-0.028$_{-0.004}^{+0.009}$~\cite{PhysRevLett.92.012004}, respectively.  
The proton asymmetries, $A_{y}^{p}$, were predicted  to be 0.01\%, 0.24\% and 0.62\% for $Q^{2}=0.13$, 0.46 and 0.97 GeV$^2$, respectively, using an estimate of the elastic intermediate state contribution in lieu of precision data~\cite{Afanasev:2002gr}. The contribution from $A_y^p$ is suppressed by the small effective proton polarization in polarized $^3$He.  The  neutron single-spin asymmetries calculated using Eq.~(\ref{eqn:pdilution}) are shown in Fig.~\ref{fig:result_n_q2} and listed in Table~\ref{tab:results_n} along with values for $f_n$. 

\begin{table*}[h]
\begin{tabular}{|c |c |c |c || c || c | c|| c| c|}
\hline
  $E$ (GeV) & $\left < E^\prime \right>$ (GeV) & $\left< \theta \right>$ deg. & $\left< Q^{2} \right>$ (GeV$^{2}$)  & $A_y^{^{\text{3}}\text{He}}(\%)$ & $f_{n}$ (Kelly) & $A_{y}^n (\%)$ (Kelly) & $f_{n}$ (Deltuva)  &$A_{y}^n (\%)$ (Deltuva)  \\
  \hline
1.245 & 1.167 & 17.0  & 0.127  & -0.126$\pm$0.015$\pm$0.027 & 0.050 & -2.92$\pm$0.36$\pm$0.64   & 0.044 & -3.32$\pm$0.40$\pm$0.72 \\
2.425 & 2.170 & 17.0 & 0.460  &-0.185$\pm$0.020$\pm$0.013 & 0.117 & -1.78$\pm$0.26$\pm$0.16    & 0.104 & -2.00$\pm$0.29$\pm$0.18 \\
3.605 & 3.070 & 17.0 & 0.967  &-0.199$\pm$0.019$\pm$0.013 & 0.159 &  -1.35$\pm$0.25$\pm$0.16  & - & -\\
    \hline
  \end{tabular}
\caption{\label{tab:results_n} Measured $^3$He asymmetries, $A_y^{^{\text{3}}\text{He}}(\%)$.  Neutron dilution factors, $f_n$, and asymmetries, $A_y^n$, were extracted using the Deltuva and Kelly models for $^3$He.  Uncertainties are statistical and systematic, respectively.}
\end{table*}

 In summary, we have reported the first measurement of the target single-spin asymmetries, $A_{y}$, from quasi-elastic $(e,e^\prime)$ scattering from a  $^3$He target polarized normal to the electron scattering plane.  This measurement demonstrates, for the first time, that the $^3$He asymmetries are clearly  non-zero and negative with a statistical significance of $(8-10)\sigma$.  Neutron asymmetries were extracted using the effective neutron polarization approximation and are also clearly  non-zero and negative.  The results are inconsistent with an estimate where only the elastic intermediate state is included~\cite{Afanasev:2002gr} but are consistent with a model using GPD input for the inelastic intermediate state contribution at $Q^2=0.97$ GeV$^2$~\cite{Chen:2004tw}.  

We acknowledge the outstanding support of the Jefferson Lab
Hall A technical staff and Accelerator Division in accomplishing this experiment. We  wish to thank Profs.~A.~Deltuva, Univ.~de Lisboa, A. Afanasev, Jefferson Lab and M. Vanderhaeghen, MAINZ,  for theoretical guidance and calculations.  This work was supported in
part by the U.S. National Science Foundation, the U.S. Department of Energy and by DOE contract DE-AC05-06OR23177, under which
Jefferson Science Associates, LLC operates the Thomas
Jefferson National Accelerator Facility, the National Science Foundation of China and UK STFC grants 57071/1, 50727/1.

\bibliography{Ay_QE_PRL}
\end{document}

%% file: author_list.tex
\author{Y.-W.~Zhang} \affiliation{Rutgers University, New Brunswick, NJ 08901, USA} \affiliation{University of Pennsylvania, Philadelphia, PA, 19104, USA}
\author{E.~Long} \affiliation{Kent State University, Kent, OH 44242, USA} 
\author{M.~Mihovilovi\v{c}} \affiliation{Jo\v{z}ef Stefan Institute, SI-1000 Ljubljana, Slovenia}
\author{G.~Jin} \affiliation{University of Virginia, Charlottesville, VA 22908, USA}
\author{K.~Allada} \affiliation{Thomas Jefferson National Accelerator Facility, Newport News, VA 23606, USA}
\author{B.~Anderson} \affiliation{Kent State University, Kent, OH 44242, USA}
\author{J.~R.~M.~Annand} \affiliation{Glasgow University, Glasgow G12 8QQ, Scotland, United Kingdom}
\author{T.~Averett} \email[Corresponding author: ]{tdaver@wm.edu} \affiliation{The College of William and Mary, Williamsburg, VA 23187, USA}
\author{W.~Boeglin} \affiliation{Florida International University, Miami, FL 33181, USA}
\author{P.~Bradshaw} \affiliation{The College of William and Mary, Williamsburg, VA 23187, USA}
\author{A.~Camsonne} \affiliation{Thomas Jefferson National Accelerator Facility, Newport News, VA 23606, USA}
\author{M.~Canan} \affiliation{Old Dominion University, Norfolk, VA 23529, USA}
\author{G.~D.~Cates} \affiliation{University of Virginia, Charlottesville, VA 22908, USA}
\author{C.~Chen} \affiliation{Hampton University, Hampton, VA 23669, USA}
\author{J.~P.~Chen} \affiliation{Thomas Jefferson National Accelerator Facility, Newport News, VA 23606, USA}
\author{E.~Chudakov} \affiliation{Thomas Jefferson National Accelerator Facility, Newport News, VA 23606, USA}
\author{R.~De~Leo} \affiliation{Universit\`a degli studi di Bari Aldo Moro, I-70121 Bari, Italy}
\author{X.~Deng} \affiliation{University of Virginia, Charlottesville, VA 22908, USA}
\author{A.~Deur} \affiliation{Thomas Jefferson National Accelerator Facility, Newport News, VA 23606, USA}
\author{C.~Dutta} \affiliation{University of Kentucky, Lexington, KY 40506, USA}
\author{L.~El~Fassi} \affiliation{Rutgers University, New Brunswick, NJ 08901, USA}
\author{D.~Flay} \affiliation{Temple University, Philadelphia, PA 19122, USA}
\author{S.~Frullani} \affiliation{Istituto Nazionale Di Fisica Nucleare, INFN/Sanita, Roma, Italy}
\author{F.~Garibaldi} \affiliation{Istituto Nazionale Di Fisica Nucleare, INFN/Sanita, Roma, Italy}
\author{H.~Gao} \affiliation{Duke University, Durham, NC 27708, USA}
\author{S.~Gilad} \affiliation{Massachusetts Institute of Technology, Cambridge, MA 02139, USA}
\author{R.~Gilman} \affiliation{Rutgers University, New Brunswick, NJ 08901, USA}
\author{O.~Glamazdin} \affiliation{Kharkov Institute of Physics and Technology, Kharkov 61108, Ukraine}
\author{S.~Golge} \affiliation{Old Dominion University, Norfolk, VA 23529, USA}
\author{J.~Gomez} \affiliation{Thomas Jefferson National Accelerator Facility, Newport News, VA 23606, USA}
\author{O.~Hansen} \affiliation{Thomas Jefferson National Accelerator Facility, Newport News, VA 23606, USA}
\author{D.~W.~Higinbotham} \affiliation{Thomas Jefferson National Accelerator Facility, Newport News, VA 23606, USA}
\author{T.~Holmstrom} \affiliation{Longwood University, Farmville, VA 23909, USA}
\author{J.~Huang} \affiliation{Massachusetts Institute of Technology, Cambridge, MA 02139, USA}\affiliation{Los Alamos National Laboratory, Los Alamos, NM 87545, USA}
\author{H.~Ibrahim} \affiliation{Cairo University, Cairo, Giza 12613, Egypt}
\author{C.~W.~de~Jager} \affiliation{Thomas Jefferson National Accelerator Facility, Newport News, VA 23606, USA}
\author{E.~Jensen} \affiliation{Christopher Newport University, Newport News VA 23606, USA}
\author{X.~Jiang} \affiliation{Los Alamos National Laboratory, Los Alamos, NM 87545, USA}
\author{J.~St.~John} \affiliation{Longwood University, Farmville, VA 23909, USA}
\author{M.~Jones} \affiliation{Thomas Jefferson National Accelerator Facility, Newport News, VA 23606, USA}
\author{H.~Kang} \affiliation{Seoul National University, Seoul, Korea}
\author{J.~Katich} \affiliation{The College of William and Mary, Williamsburg, VA 23187, USA}
\author{H.~P.~Khanal} \affiliation{Florida International University, Miami, FL 33181, USA}
\author{P.~King} \affiliation{Ohio University, Athens, OH 45701, USA}
\author{W.~Korsch} \affiliation{University of Kentucky, Lexington, KY 40506, USA}
\author{J.~LeRose} \affiliation{Thomas Jefferson National Accelerator Facility, Newport News, VA 23606, USA}
\author{R.~Lindgren} \affiliation{University of Virginia, Charlottesville, VA 22908, USA}
\author{H.-J.~Lu} \affiliation{Huangshan University, People's Republic of China}
\author{W.~Luo} \affiliation{Lanzhou University, Lanzhou, Gansu, 730000, People's Republic of China}
\author{P.~Markowitz} \affiliation{Florida International University, Miami, FL 33181, USA}
\author{M.~Meziane} \affiliation{The College of William and Mary, Williamsburg, VA 23187, USA}
\author{R.~Michaels} \affiliation{Thomas Jefferson National Accelerator Facility, Newport News, VA 23606, USA}
\author{B.~Moffit} \affiliation{Thomas Jefferson National Accelerator Facility, Newport News, VA 23606, USA}
\author{P.~Monaghan} \affiliation{Hampton University, Hampton, VA 23669, USA}
\author{N.~Muangma} \affiliation{Massachusetts Institute of Technology, Cambridge, MA 02139, USA}
\author{S.~Nanda} \affiliation{Thomas Jefferson National Accelerator Facility, Newport News, VA 23606, USA}
\author{B.~E.~Norum} \affiliation{University of Virginia, Charlottesville, VA 22908, USA}
\author{K.~Pan} \affiliation{Massachusetts Institute of Technology, Cambridge, MA 02139, USA}
\author{D.~Parno} \affiliation{Carnegie Mellon University, Pittsburgh, PA 15213, USA}
\author{E.~Piasetzky} \affiliation{Tel Aviv University, Tel Aviv 69978, Israel}
\author{M.~Posik} \affiliation{Temple University, Philadelphia, PA 19122, USA}
\author{V.~Punjabi} \affiliation{Norfolk State University, Norfolk, VA 23504, USA}
\author{A.~J.~R.~Puckett} \affiliation{Los Alamos National Laboratory, Los Alamos, NM 87545, USA}
\author{X.~Qian} \affiliation{Duke University, Durham, NC 27708, USA}
\author{Y.~Qiang} \affiliation{Thomas Jefferson National Accelerator Facility, Newport News, VA 23606, USA}
\author{X.~Qiu} \affiliation{Lanzhou University, Lanzhou, Gansu, 730000, People's Republic of China}
\author{S.~Riordan} \affiliation{University of Virginia, Charlottesville, VA 22908, USA}
\author{G.~Ron} \affiliation{Hebrew University of Jerusalem, Jerusalem 91904 Israel}
\author{A.~Saha} \email[Deceased.]{} \affiliation{Thomas Jefferson National Accelerator Facility, Newport News, VA 23606, USA}
\author{B.~Sawatzky} \affiliation{Thomas Jefferson National Accelerator Facility, Newport News, VA 23606, USA}
\author{R.~Schiavilla} \affiliation{Thomas Jefferson National Accelerator Facility, Newport News, VA 23606, USA} \affiliation{Old Dominion University, Norfolk, VA 23529, USA}
\author{B.~Schoenrock} \affiliation{Northern Michigan University, Marquette, MI 49855, USA}
\author{M.~Shabestari} \affiliation{University of Virginia, Charlottesville, VA 22908, USA}
\author{A.~Shahinyan} \affiliation{Yerevan Physics Institute, Yerevan, Armenia}
\author{S.~\v{S}irca} \affiliation{University of Ljubljana, SI-1000 Ljubljana, Slovenia}\affiliation{Jo\v{z}ef Stefan Institute, SI-1000 Ljubljana, Slovenia}
\author{R.~Subedi} \affiliation{George Washington University, Washington, D.C. 20052, USA}
\author{V.~Sulkosky} \affiliation{Massachusetts Institute of Technology, Cambridge, MA 02139, USA}
\author{W.~A.~Tobias} \affiliation{University of Virginia, Charlottesville, VA 22908, USA}
\author{W.~Tireman} \affiliation{Northern Michigan University, Marquette, MI 49855, USA}
\author{G.~M.~Urciuoli} \affiliation{Istituto Nazionale Di Fisica Nucleare, INFN/Sanita, Roma, Italy}
\author{D.~Wang} \affiliation{University of Virginia, Charlottesville, VA 22908, USA}
\author{K.~Wang} \affiliation{University of Virginia, Charlottesville, VA 22908, USA}
\author{Y.~Wang} \affiliation{University of Illinois at Urbana-Champaign, Urbana, IL 61801, USA}
\author{J.~Watson} \affiliation{Thomas Jefferson National Accelerator Facility, Newport News, VA 23606, USA}
\author{B.~Wojtsekhowski} \affiliation{Thomas Jefferson National Accelerator Facility, Newport News, VA 23606, USA}
\author{Y.~Ye}\affiliation{University of Science and Technology, Hefei, People's Republic of China}
\author{Z.~Ye} \affiliation{Hampton University, Hampton, VA 23669, USA}
\author{X.~Zhan} \affiliation{Massachusetts Institute of Technology, Cambridge, MA 02139, USA}
\author{Y.~Zhang} \affiliation{Lanzhou University, Lanzhou, Gansu, 730000, People's Republic of China}
\author{X.~Zheng} \affiliation{University of Virginia, Charlottesville, VA 22908, USA}
\author{B.~Zhao} \affiliation{The College of William and Mary, Williamsburg, VA 23187, USA}
\author{L.~Zhu} \affiliation{Hampton University, Hampton, VA 23669, USA}
\collaboration{The Jefferson Lab Hall A Collaboration} \noaffiliation

%% file: Ay_QE_PRL.bbl
\begin{thebibliography}{26}%
\makeatletter
\providecommand \@ifxundefined [1]{%
 \@ifx{#1\undefined}
}%
\providecommand \@ifnum [1]{%
 \ifnum #1\expandafter \@firstoftwo
 \else \expandafter \@secondoftwo
 \fi
}%
\providecommand \@ifx [1]{%
 \ifx #1\expandafter \@firstoftwo
 \else \expandafter \@secondoftwo
 \fi
}%
\providecommand \natexlab [1]{#1}%
\providecommand \enquote  [1]{``#1''}%
\providecommand \bibnamefont  [1]{#1}%
\providecommand \bibfnamefont [1]{#1}%
\providecommand \citenamefont [1]{#1}%
\providecommand \href@noop [0]{\@secondoftwo}%
\providecommand \href [0]{\begingroup \@sanitize@url \@href}%
\providecommand \@href[1]{\@@startlink{#1}\@@href}%
\providecommand \@@href[1]{\endgroup#1\@@endlink}%
\providecommand \@sanitize@url [0]{\catcode `\\12\catcode `\$12\catcode
  `\&12\catcode `\#12\catcode `\^12\catcode `\_12\catcode `\%12\relax}%
\providecommand \@@startlink[1]{}%
\providecommand \@@endlink[0]{}%
\providecommand \url  [0]{\begingroup\@sanitize@url \@url }%
\providecommand \@url [1]{\endgroup\@href {#1}{\urlprefix }}%
\providecommand \urlprefix  [0]{URL }%
\providecommand \Eprint [0]{\href }%
\providecommand \doibase [0]{http://dx.doi.org/}%
\providecommand \selectlanguage [0]{\@gobble}%
\providecommand \bibinfo  [0]{\@secondoftwo}%
\providecommand \bibfield  [0]{\@secondoftwo}%
\providecommand \translation [1]{[#1]}%
\providecommand \BibitemOpen [0]{}%
\providecommand \bibitemStop [0]{}%
\providecommand \bibitemNoStop [0]{.\EOS\space}%
\providecommand \EOS [0]{\spacefactor3000\relax}%
\providecommand \BibitemShut  [1]{\csname bibitem#1\endcsname}%
\let\auto@bib@innerbib\@empty
\bibitem [{\citenamefont {Chen}\ \emph {et~al.}(2004)\citenamefont {Chen},
  \citenamefont {Afanasev}, \citenamefont {Brodsky}, \citenamefont {Carlson},\
  and\ \citenamefont {Vanderhaeghen}}]{Chen:2004tw}%
  \BibitemOpen
  \bibfield  {author} {\bibinfo {author} {\bibfnamefont {Y.}~\bibnamefont
  {Chen}}, \bibinfo {author} {\bibfnamefont {A.}~\bibnamefont {Afanasev}},
  \bibinfo {author} {\bibfnamefont {S.}~\bibnamefont {Brodsky}}, \bibinfo
  {author} {\bibfnamefont {C.}~\bibnamefont {Carlson}}, \ and\ \bibinfo
  {author} {\bibfnamefont {M.}~\bibnamefont {Vanderhaeghen}},\ }\href {\doibase
  10.1103/PhysRevLett.93.122301} {\bibfield  {journal} {\bibinfo  {journal}
  {Phys.Rev.Lett.}\ }\textbf {\bibinfo {volume} {93}},\ \bibinfo {pages}
  {122301} (\bibinfo {year} {2004})}\BibitemShut {NoStop}%
\bibitem [{\citenamefont {Katich}\ \emph {et~al.}(2014)\citenamefont {Katich},
  \citenamefont {Qian}, \citenamefont {Zhao}, \citenamefont {Allada},
  \citenamefont {Aniol} \emph {et~al.}}]{Katich:2013atq}%
  \BibitemOpen
  \bibfield  {author} {\bibinfo {author} {\bibfnamefont {J.}~\bibnamefont
  {Katich}}, \bibinfo {author} {\bibfnamefont {X.}~\bibnamefont {Qian}},
  \bibinfo {author} {\bibfnamefont {Y.}~\bibnamefont {Zhao}}, \bibinfo {author}
  {\bibfnamefont {K.}~\bibnamefont {Allada}}, \bibinfo {author} {\bibfnamefont
  {K.}~\bibnamefont {Aniol}},  \emph {et~al.},\ }\href {\doibase
  10.1103/PhysRevLett.113.022502} {\bibfield  {journal} {\bibinfo  {journal}
  {Phys.Rev.Lett.}\ }\textbf {\bibinfo {volume} {113}},\ \bibinfo {pages}
  {022502} (\bibinfo {year} {2014})},\ \Eprint {http://arxiv.org/abs/1311.0197}
  {arXiv:1311.0197 [nucl-ex]} \BibitemShut {NoStop}%
\bibitem [{\citenamefont {Arrington}(2003)}]{g63}%
  \BibitemOpen
  \bibfield  {author} {\bibinfo {author} {\bibfnamefont {J.}~\bibnamefont
  {Arrington}},\ }\href {\doibase 10.1103/PhysRevC.68.034325} {\bibfield
  {journal} {\bibinfo  {journal} {Phys. Rev. C}\ }\textbf {\bibinfo {volume}
  {68}},\ \bibinfo {pages} {034325} (\bibinfo {year} {2003})}\BibitemShut
  {NoStop}%
\bibitem [{\citenamefont {Arrington}\ \emph {et~al.}(2011)\citenamefont
  {Arrington}, \citenamefont {Blunden},\ and\ \citenamefont
  {Melnitchouk}}]{Arrington:2011dn}%
  \BibitemOpen
  \bibfield  {author} {\bibinfo {author} {\bibfnamefont {J.}~\bibnamefont
  {Arrington}}, \bibinfo {author} {\bibfnamefont {P.}~\bibnamefont {Blunden}},
  \ and\ \bibinfo {author} {\bibfnamefont {W.}~\bibnamefont {Melnitchouk}},\
  }\href {\doibase 10.1016/j.ppnp.2011.07.003} {\bibfield  {journal} {\bibinfo
  {journal} {Prog.Part.Nucl.Phys.}\ }\textbf {\bibinfo {volume} {66}},\
  \bibinfo {pages} {782} (\bibinfo {year} {2011})},\ \Eprint
  {http://arxiv.org/abs/1105.0951} {arXiv:1105.0951 [nucl-th]} \BibitemShut
  {NoStop}%
\bibitem [{\citenamefont {Puckett}\ \emph {et~al.}(2012)\citenamefont
  {Puckett}, \citenamefont {Brash}, \citenamefont {Gayou}, \citenamefont
  {Jones}, \citenamefont {Pentchev} \emph {et~al.}}]{Puckett:2011xg}%
  \BibitemOpen
  \bibfield  {author} {\bibinfo {author} {\bibfnamefont {A.}~\bibnamefont
  {Puckett}}, \bibinfo {author} {\bibfnamefont {E.}~\bibnamefont {Brash}},
  \bibinfo {author} {\bibfnamefont {O.}~\bibnamefont {Gayou}}, \bibinfo
  {author} {\bibfnamefont {M.}~\bibnamefont {Jones}}, \bibinfo {author}
  {\bibfnamefont {L.}~\bibnamefont {Pentchev}},  \emph {et~al.},\ }\href
  {\doibase 10.1103/PhysRevC.85.045203} {\bibfield  {journal} {\bibinfo
  {journal} {Phys.Rev.}\ }\textbf {\bibinfo {volume} {C85}},\ \bibinfo {pages}
  {045203} (\bibinfo {year} {2012})}\BibitemShut {NoStop}%
\bibitem [{\citenamefont {Blunden}\ \emph {et~al.}(2003)\citenamefont
  {Blunden}, \citenamefont {Melnitchouk},\ and\ \citenamefont
  {Tjon}}]{Blunden_GeGm}%
  \BibitemOpen
  \bibfield  {author} {\bibinfo {author} {\bibfnamefont {P.~G.}\ \bibnamefont
  {Blunden}}, \bibinfo {author} {\bibfnamefont {W.}~\bibnamefont
  {Melnitchouk}}, \ and\ \bibinfo {author} {\bibfnamefont {J.~A.}\ \bibnamefont
  {Tjon}},\ }\href {\doibase 10.1103/PhysRevLett.91.142304} {\bibfield
  {journal} {\bibinfo  {journal} {Phys. Rev. Lett.}\ }\textbf {\bibinfo
  {volume} {91}},\ \bibinfo {pages} {142304} (\bibinfo {year}
  {2003})}\BibitemShut {NoStop}%
\bibitem [{\citenamefont {{Carlson}}\ and\ \citenamefont
  {{Vanderhaeghen}}(2007)}]{carl_2gamma}%
  \BibitemOpen
  \bibfield  {author} {\bibinfo {author} {\bibfnamefont {C.~E.}\ \bibnamefont
  {{Carlson}}}\ and\ \bibinfo {author} {\bibfnamefont {M.}~\bibnamefont
  {{Vanderhaeghen}}},\ }\href {\doibase 10.1146/annurev.nucl.57.090506.123116}
  {\bibfield  {journal} {\bibinfo  {journal} {Annu. Rev. Nucl. Part. Sci.}\
  }\textbf {\bibinfo {volume} {57}},\ \bibinfo {pages} {171} (\bibinfo {year}
  {2007})}\BibitemShut {NoStop}%
\bibitem [{\citenamefont {Christ}\ and\ \citenamefont
  {Lee}(1966)}]{Christ_Lee}%
  \BibitemOpen
  \bibfield  {author} {\bibinfo {author} {\bibfnamefont {N.}~\bibnamefont
  {Christ}}\ and\ \bibinfo {author} {\bibfnamefont {T.~D.}\ \bibnamefont
  {Lee}},\ }\href@noop {} {\bibfield  {journal} {\bibinfo  {journal} {Phys.
  Rev.}\ }\textbf {\bibinfo {volume} {143}},\ \bibinfo {pages} {1310} (\bibinfo
  {year} {1966,})}\BibitemShut {NoStop}%
\bibitem [{\citenamefont {Cahn}\ and\ \citenamefont {Tsai}(1970)}]{Cahn_2g}%
  \BibitemOpen
  \bibfield  {author} {\bibinfo {author} {\bibfnamefont {R.~N.}\ \bibnamefont
  {Cahn}}\ and\ \bibinfo {author} {\bibfnamefont {Y.~S.}\ \bibnamefont
  {Tsai}},\ }\href {\doibase 10.1103/PhysRevD.2.870} {\bibfield  {journal}
  {\bibinfo  {journal} {Phys. Rev.}\ }\textbf {\bibinfo {volume} {{\bf D2}}},\
  \bibinfo {pages} {870} (\bibinfo {year} {1970})}\BibitemShut {NoStop}%
\bibitem [{\citenamefont {Afanasev}\ \emph {et~al.}(2008)\citenamefont
  {Afanasev}, \citenamefont {Strikman},\ and\ \citenamefont
  {Weiss}}]{Afanasev_DIS}%
  \BibitemOpen
  \bibfield  {author} {\bibinfo {author} {\bibfnamefont {A.}~\bibnamefont
  {Afanasev}}, \bibinfo {author} {\bibfnamefont {M.}~\bibnamefont {Strikman}},
  \ and\ \bibinfo {author} {\bibfnamefont {C.}~\bibnamefont {Weiss}},\ }\href
  {\doibase 10.1103/PhysRevD.77.014028} {\bibfield  {journal} {\bibinfo
  {journal} {Phys. Rev.}\ }\textbf {\bibinfo {volume} {{\bf D77}}},\ \bibinfo
  {pages} {014028} (\bibinfo {year} {2008})}\BibitemShut {NoStop}%
\bibitem [{\citenamefont {Pasquini}\ and\ \citenamefont
  {Vanderhaeghen}(2004)}]{Pasquini:2004pv}%
  \BibitemOpen
  \bibfield  {author} {\bibinfo {author} {\bibfnamefont {B.}~\bibnamefont
  {Pasquini}}\ and\ \bibinfo {author} {\bibfnamefont {M.}~\bibnamefont
  {Vanderhaeghen}},\ }\href {\doibase 10.1103/PhysRevC.70.045206} {\bibfield
  {journal} {\bibinfo  {journal} {Phys.Rev.}\ }\textbf {\bibinfo {volume}
  {C70}},\ \bibinfo {pages} {045206} (\bibinfo {year} {2004})},\ \Eprint
  {http://arxiv.org/abs/hep-ph/0405303} {arXiv:hep-ph/0405303 [hep-ph]}
  \BibitemShut {NoStop}%
\bibitem [{\citenamefont {Powell}\ \emph {et~al.}(1970)\citenamefont {Powell},
  \citenamefont {Borghini}, \citenamefont {Chamberlain}, \citenamefont
  {Fuzesy}, \citenamefont {Morehouse} \emph {et~al.}}]{Powell:1970qt}%
  \BibitemOpen
  \bibfield  {author} {\bibinfo {author} {\bibfnamefont {T.}~\bibnamefont
  {Powell}}, \bibinfo {author} {\bibfnamefont {M.}~\bibnamefont {Borghini}},
  \bibinfo {author} {\bibfnamefont {O.}~\bibnamefont {Chamberlain}}, \bibinfo
  {author} {\bibfnamefont {R.~Z.}\ \bibnamefont {Fuzesy}}, \bibinfo {author}
  {\bibfnamefont {C.~C.}\ \bibnamefont {Morehouse}},  \emph {et~al.},\ }\href
  {\doibase 10.1103/PhysRevLett.24.753} {\bibfield  {journal} {\bibinfo
  {journal} {Phys.Rev.Lett.}\ }\textbf {\bibinfo {volume} {24}},\ \bibinfo
  {pages} {753} (\bibinfo {year} {1970})}\BibitemShut {NoStop}%
\bibitem [{\citenamefont {Kuhn}\ \emph {et~al.}(2009)\citenamefont {Kuhn},
  \citenamefont {Chen},\ and\ \citenamefont {Leader}}]{Kuhn:2008sy}%
  \BibitemOpen
  \bibfield  {author} {\bibinfo {author} {\bibfnamefont {S.}~\bibnamefont
  {Kuhn}}, \bibinfo {author} {\bibfnamefont {J.-P.}\ \bibnamefont {Chen}}, \
  and\ \bibinfo {author} {\bibfnamefont {E.}~\bibnamefont {Leader}},\ }\href
  {\doibase 10.1016/j.ppnp.2009.02.001} {\bibfield  {journal} {\bibinfo
  {journal} {Prog.Part.Nucl.Phys.}\ }\textbf {\bibinfo {volume} {63}},\
  \bibinfo {pages} {1} (\bibinfo {year} {2009})},\ \Eprint
  {http://arxiv.org/abs/0812.3535} {arXiv:0812.3535 [hep-ph]} \BibitemShut
  {NoStop}%
\bibitem [{\citenamefont {Qian}\ \emph {et~al.}(2011)\citenamefont {Qian} \emph
  {et~al.}}]{Qian:2011py}%
  \BibitemOpen
  \bibfield  {author} {\bibinfo {author} {\bibfnamefont {X.}~\bibnamefont
  {Qian}} \emph {et~al.} (\bibinfo {collaboration} {Jefferson Lab Hall A
  Collaboration}),\ }\href {\doibase 10.1103/PhysRevLett.107.072003} {\bibfield
   {journal} {\bibinfo  {journal} {Phys.Rev.Lett.}\ }\textbf {\bibinfo {volume}
  {107}},\ \bibinfo {pages} {072003} (\bibinfo {year} {2011})},\ \Eprint
  {http://arxiv.org/abs/1106.0363} {arXiv:1106.0363 [nucl-ex]} \BibitemShut
  {NoStop}%
\bibitem [{\citenamefont {Bissey}\ \emph {et~al.}(2002)\citenamefont {Bissey},
  \citenamefont {Guzey}, \citenamefont {Strikman},\ and\ \citenamefont
  {Thomas}}]{PhysRevC.65.064317}%
  \BibitemOpen
  \bibfield  {author} {\bibinfo {author} {\bibfnamefont {F.}~\bibnamefont
  {Bissey}}, \bibinfo {author} {\bibfnamefont {V.}~\bibnamefont {Guzey}},
  \bibinfo {author} {\bibfnamefont {M.}~\bibnamefont {Strikman}}, \ and\
  \bibinfo {author} {\bibfnamefont {A.}~\bibnamefont {Thomas}},\ }\href
  {\doibase 10.1103/PhysRevC.65.064317} {\bibfield  {journal} {\bibinfo
  {journal} {Phys. Rev. C}\ }\textbf {\bibinfo {volume} {65}},\ \bibinfo
  {pages} {064317} (\bibinfo {year} {2002})}\BibitemShut {NoStop}%
\bibitem [{\citenamefont {Singh}\ \emph {et~al.}(2013)\citenamefont {Singh},
  \citenamefont {Dolph}, \citenamefont {Tobias}, \citenamefont {Averett},
  \citenamefont {Kelleher} \emph {et~al.}}]{Singh:2013nja}%
  \BibitemOpen
  \bibfield  {author} {\bibinfo {author} {\bibfnamefont {J.}~\bibnamefont
  {Singh}}, \bibinfo {author} {\bibfnamefont {P.}~\bibnamefont {Dolph}},
  \bibinfo {author} {\bibfnamefont {W.}~\bibnamefont {Tobias}}, \bibinfo
  {author} {\bibfnamefont {T.}~\bibnamefont {Averett}}, \bibinfo {author}
  {\bibfnamefont {A.}~\bibnamefont {Kelleher}},  \emph {et~al.},\ }\href@noop
  {} {\  (\bibinfo {year} {2013})},\ \Eprint {http://arxiv.org/abs/1309.4004}
  {arXiv:1309.4004 [physics.atom-ph]} \BibitemShut {NoStop}%
\bibitem [{\citenamefont {Romalis}\ and\ \citenamefont
  {Cates}(1998)}]{Romalis_EPR}%
  \BibitemOpen
  \bibfield  {author} {\bibinfo {author} {\bibfnamefont {M.~V.}\ \bibnamefont
  {Romalis}}\ and\ \bibinfo {author} {\bibfnamefont {G.~D.}\ \bibnamefont
  {Cates}},\ }\href {\doibase 10.1103/PhysRevA.58.3004} {\bibfield  {journal}
  {\bibinfo  {journal} {Phys. Rev. A}\ }\textbf {\bibinfo {volume} {58}},\
  \bibinfo {pages} {3004} (\bibinfo {year} {1998})}\BibitemShut {NoStop}%
\bibitem [{\citenamefont {Alcorn}\ \emph {et~al.}(2004)\citenamefont {Alcorn},
  \citenamefont {Anderson}, \citenamefont {Aniol}, \citenamefont {Annand},
  \citenamefont {Auerbach} \emph {et~al.}}]{Alcorn:2004sb}%
  \BibitemOpen
  \bibfield  {author} {\bibinfo {author} {\bibfnamefont {J.}~\bibnamefont
  {Alcorn}}, \bibinfo {author} {\bibfnamefont {B.}~\bibnamefont {Anderson}},
  \bibinfo {author} {\bibfnamefont {K.}~\bibnamefont {Aniol}}, \bibinfo
  {author} {\bibfnamefont {J.}~\bibnamefont {Annand}}, \bibinfo {author}
  {\bibfnamefont {L.}~\bibnamefont {Auerbach}},  \emph {et~al.},\ }\href
  {\doibase 10.1016/j.nima.2003.11.415} {\bibfield  {journal} {\bibinfo
  {journal} {Nucl.Instrum.Meth.}\ }\textbf {\bibinfo {volume} {A522}},\
  \bibinfo {pages} {294} (\bibinfo {year} {2004})}\BibitemShut {NoStop}%
\bibitem [{\citenamefont {Scopetta}(2007)}]{Scopetta}%
  \BibitemOpen
  \bibfield  {author} {\bibinfo {author} {\bibfnamefont {S.}~\bibnamefont
  {Scopetta}},\ }\href@noop {} {\bibfield  {journal} {\bibinfo  {journal}
  {Phys. Rev. D}\ }\textbf {\bibinfo {volume} {75}},\ \bibinfo {pages} {054005}
  (\bibinfo {year} {2007})}\BibitemShut {NoStop}%
\bibitem [{\citenamefont {Kelly}(2004)}]{PhysRevC.70.068202}%
  \BibitemOpen
  \bibfield  {author} {\bibinfo {author} {\bibfnamefont {J.~J.}\ \bibnamefont
  {Kelly}},\ }\href {\doibase 10.1103/PhysRevC.70.068202} {\bibfield  {journal}
  {\bibinfo  {journal} {Phys. Rev. C}\ }\textbf {\bibinfo {volume} {70}},\
  \bibinfo {pages} {068202} (\bibinfo {year} {2004})}\BibitemShut {NoStop}%
\bibitem [{\citenamefont {Deltuva}\ \emph {et~al.}(2003)\citenamefont
  {Deltuva}, \citenamefont {Machleidt},\ and\ \citenamefont
  {Sauer}}]{PhysRevC.68.024005}%
  \BibitemOpen
  \bibfield  {author} {\bibinfo {author} {\bibfnamefont {A.}~\bibnamefont
  {Deltuva}}, \bibinfo {author} {\bibfnamefont {R.}~\bibnamefont {Machleidt}},
  \ and\ \bibinfo {author} {\bibfnamefont {P.}~\bibnamefont {Sauer}},\ }\href
  {\doibase 10.1103/PhysRevC.68.024005} {\bibfield  {journal} {\bibinfo
  {journal} {Phys. Rev. C}\ }\textbf {\bibinfo {volume} {68}},\ \bibinfo
  {pages} {024005} (\bibinfo {year} {2003})}\BibitemShut {NoStop}%
\bibitem [{\citenamefont {Deltuva}\ \emph
  {et~al.}(2004{\natexlab{a}})\citenamefont {Deltuva}, \citenamefont {Yuan},
  \citenamefont {Adam}, \citenamefont {Fonseca},\ and\ \citenamefont
  {Sauer}}]{PhysRevC.69.034004}%
  \BibitemOpen
  \bibfield  {author} {\bibinfo {author} {\bibfnamefont {A.}~\bibnamefont
  {Deltuva}}, \bibinfo {author} {\bibfnamefont {L.}~\bibnamefont {Yuan}},
  \bibinfo {author} {\bibfnamefont {J.}~\bibnamefont {Adam}}, \bibinfo {author}
  {\bibfnamefont {A.}~\bibnamefont {Fonseca}}, \ and\ \bibinfo {author}
  {\bibfnamefont {P.}~\bibnamefont {Sauer}},\ }\href {\doibase
  10.1103/PhysRevC.69.034004} {\bibfield  {journal} {\bibinfo  {journal} {Phys.
  Rev. C}\ }\textbf {\bibinfo {volume} {69}},\ \bibinfo {pages} {034004}
  (\bibinfo {year} {2004}{\natexlab{a}})}\BibitemShut {NoStop}%
\bibitem [{\citenamefont {Deltuva}\ \emph
  {et~al.}(2004{\natexlab{b}})\citenamefont {Deltuva}, \citenamefont {Yuan},
  \citenamefont {Adam},\ and\ \citenamefont {Sauer}}]{PhysRevC.70.034004}%
  \BibitemOpen
  \bibfield  {author} {\bibinfo {author} {\bibfnamefont {A.}~\bibnamefont
  {Deltuva}}, \bibinfo {author} {\bibfnamefont {L.}~\bibnamefont {Yuan}},
  \bibinfo {author} {\bibfnamefont {J.}~\bibnamefont {Adam}}, \ and\ \bibinfo
  {author} {\bibfnamefont {P.}~\bibnamefont {Sauer}},\ }\href {\doibase
  10.1103/PhysRevC.70.034004} {\bibfield  {journal} {\bibinfo  {journal} {Phys.
  Rev. C}\ }\textbf {\bibinfo {volume} {70}},\ \bibinfo {pages} {034004}
  (\bibinfo {year} {2004}{\natexlab{b}})}\BibitemShut {NoStop}%
\bibitem [{\citenamefont {Deltuva}\ \emph {et~al.}(2005)\citenamefont
  {Deltuva}, \citenamefont {Fonseca},\ and\ \citenamefont
  {Sauer}}]{PhysRevC.72.054004}%
  \BibitemOpen
  \bibfield  {author} {\bibinfo {author} {\bibfnamefont {A.}~\bibnamefont
  {Deltuva}}, \bibinfo {author} {\bibfnamefont {A.}~\bibnamefont {Fonseca}}, \
  and\ \bibinfo {author} {\bibfnamefont {P.}~\bibnamefont {Sauer}},\ }\href
  {\doibase 10.1103/PhysRevC.72.054004} {\bibfield  {journal} {\bibinfo
  {journal} {Phys. Rev. C}\ }\textbf {\bibinfo {volume} {72}},\ \bibinfo
  {pages} {054004} (\bibinfo {year} {2005})}\BibitemShut {NoStop}%
\bibitem [{\citenamefont {Zheng}\ \emph {et~al.}(2004)\citenamefont {Zheng}
  \emph {et~al.}}]{PhysRevLett.92.012004}%
  \BibitemOpen
  \bibfield  {author} {\bibinfo {author} {\bibfnamefont {X.}~\bibnamefont
  {Zheng}} \emph {et~al.} (\bibinfo {collaboration} {Jefferson Lab Hall A
  Collaboration}),\ }\href {\doibase 10.1103/PhysRevLett.92.012004} {\bibfield
  {journal} {\bibinfo  {journal} {Phys. Rev. Lett.}\ }\textbf {\bibinfo
  {volume} {92}},\ \bibinfo {pages} {012004} (\bibinfo {year}
  {2004})}\BibitemShut {NoStop}%
\bibitem [{\citenamefont {Afanasev}\ \emph {et~al.}(2002)\citenamefont
  {Afanasev}, \citenamefont {Akushevich},\ and\ \citenamefont
  {Merenkov}}]{Afanasev:2002gr}%
  \BibitemOpen
  \bibfield  {author} {\bibinfo {author} {\bibfnamefont {A.}~\bibnamefont
  {Afanasev}}, \bibinfo {author} {\bibfnamefont {I.}~\bibnamefont
  {Akushevich}}, \ and\ \bibinfo {author} {\bibfnamefont {N.}~\bibnamefont
  {Merenkov}},\ }\href@noop {} {\  (\bibinfo {year} {2002})},\ \Eprint
  {http://arxiv.org/abs/hep-ph/0208260} {arXiv:hep-ph/0208260 [hep-ph]}
  \BibitemShut {NoStop}%
\end{thebibliography}%
